\documentclass[aps, twocolumn,prl,superscriptaddress]{revtex4-2}
\usepackage[utf8]{inputenc}
\usepackage{times}
\usepackage{graphicx}
\usepackage[usenames,dvipsnames]{xcolor}
\usepackage{amsmath}
\usepackage{amssymb}
\usepackage{bm}
\usepackage{hyperref}
\usepackage{siunitx}
\usepackage{listings}

\DeclareMathOperator{\tr}{tr}

\begin{document}

\title{Anomalous phonon dispersion near yielding in athermal crystals}

\author{Fumiaki Nakai}
\email{fumiaki.nakai@ess.sci.osaka-u.ac.jp}
\affiliation{Department of Earth and Space Science, The University of Osaka, 1-1 Machikaneyama, Toyonaka, Osaka 560-0043, Japan}
\author{Michio Otsuki}
\affiliation{Institute of Science and Engineering, Shimane University, 1060 Nishikawatsu-cho, Matsue, Shimane 690-8504, Japan}
\author{Kuniyasu Saitoh}
\affiliation{Department of Physics, Faculty of Science, Kyoto Sangyo University, Motoyama, Kamigamo, Kita-ku, Kyoto 603-8555, Japan}
\author{Hiroaki Katsuragi}
\affiliation{Department of Earth and Space Science, The University of Osaka, 1-1 Machikaneyama, Toyonaka, Osaka 560-0043, Japan}

\begin{abstract}
Vibrational properties of ordered athermal solids near yielding remain poorly understood. We show that yielding in a sheared crystal is governed not by a single localized instability but by directionally extended multimode softening that forms a cross-shaped low-frequency region in wave number space. Near yielding, the acoustic dispersion $\omega\sim k$ is replaced by $\omega\sim k^2$ along the soft direction, and the vibrational density of states crosses over from Debye to non-Debye scaling, with a diverging length scale. We analytically derive these scaling laws.
\end{abstract}

\maketitle

{\it Introduction---}
Yielding in crystalline solids composed of athermal particles bridges microscopic crystal physics~\cite{Kittel2004-zm} and macroscopic granular mechanics~\cite{Andreotti2013-yo}, making it a subject of broad interest across materials science~\cite{Bragg1947-ua, Bragg1949-km, Karuriya2023-me, Karuriya2024-cb, Nakai2025-lk, Kool2022-qp, Daniels2005-wn}. Since Bragg's bubble raft experiments~\cite{Bragg1947-ua, Bragg1949-km}, it has been recognized that prototypical mechanisms of crystalline plasticity—distinct from those in amorphous solids—emerge in athermal particulate media. More recent studies~\cite{Karuriya2023-me, Nakai2025-lk} show that the yielding mechanism and strength depend sensitively on crystal structure, which suggests that structural order can serve as a design parameter for controlling mechanical properties. While the stability of crystalline solids is conventionally assessed through the positive definiteness of the elastic stiffness tensor~\cite{Mouhat2014-cx}, a systematic connection between mechanical instability in athermal crystals and the vibrational spectrum---such as phonon dispersion relations or the vibrational density of states~\cite{Kittel2004-zm, Mizuno2017-gh}---has remained elusive.

Mechanical stability in athermal solids can be formulated in terms of the Hessian matrix, defined as the curvature of the potential energy landscape~\cite{Maloney2004-wh, Maloney2006-xw, Andreotti2013-yo, Petit2025-vz}, with yielding occurring when the smallest eigenvalue of the Hessian vanishes. Within this framework, considerable progress has been made for athermal amorphous solids.
Near yielding, the lowest eigenvalue exhibits power-law scaling, accompanied by the emergence of spatially localized unstable modes~\cite{Maloney2004-wh, Maloney2006-xw}. These localized instabilities are predicted to produce characteristic low-frequency scaling of the vibrational density of states~\cite{Xu2017-dt, Tanguy2010-fi, Krishnan2022-oi}.
A picture is thus emerging in which yielding is governed by a localized soft mode~\cite{Nicolas2018-zj}.
In contrast, despite numerous studies on yielding in athermal crystalline media~\cite{Bragg1947-ua, Bragg1949-km, Karuriya2023-me, Nakai2025-lk, Otsuki2023-vx}, it remains unclear how mechanical instability develops in structurally ordered systems near yielding. What is the nature of the softening that precedes instability in athermal crystals?

\begin{figure}
    \centering
    \includegraphics[width=0.8\linewidth]{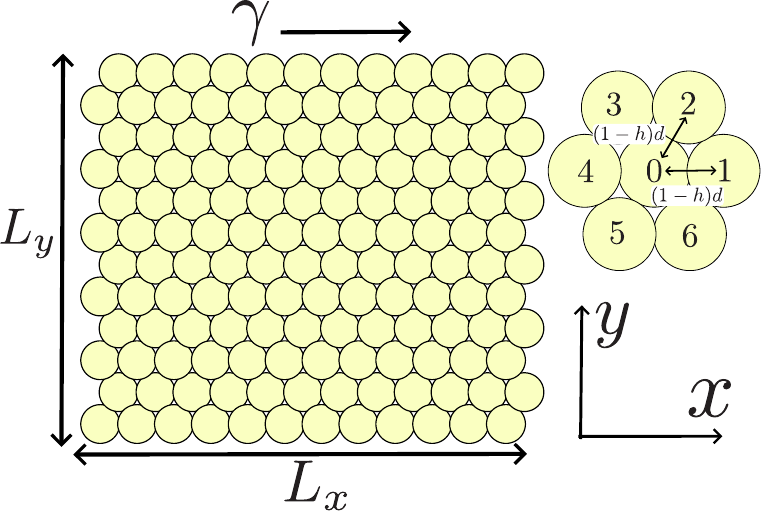}
    \caption{
    Schematic of a two-dimensional crystalline system composed of $N_x \times N_y$ Hertzian particles of diameter $d$ arranged on a triangular lattice. 
    The initial center-to-center distance between neighboring particles is $(1-h)d$, where $h$ denotes the overlap parameter. 
    Periodic boundary conditions are imposed in both the $x$ and $y$ directions.
    A quasistatic shear strain $\gamma$ is applied, and yielding occurs at a characteristic strain $\gamma_c$, defined as the point where the smallest eigenvalue of the Hessian first vanishes.
    In the theoretical analysis, a representative particle labeled $0$ and its six nearest neighbors labeled $1$–$6$ are explicitly considered.
}
    \label{figmain:setup}
\end{figure}

In this Letter, we investigate the mechanical properties of a perfect crystal composed of Hertzian particles under shear near yielding. Owing to crystalline symmetry, the eigenvalues and eigenvectors of the Hessian can be expressed as functions of the wave number~\cite{Merkel2010-kz, Merkel2011-xv, Pichard2016-ad}, allowing a detailed analysis of the vibrational spectrum as reflected in the dispersion relation and the vibrational density of states. We find that, near yielding, multiple vibrational modes soften simultaneously along specific directions in wave number space. As a consequence of this multimode softening, the conventional acoustic dispersion at long wavelengths $\omega \sim k$ is replaced by $\omega \sim k^{2}$. Correspondingly, the vibrational density of states changes from Debye scaling, $D(\omega) \sim \omega$, to non-Debye scaling, $D(\omega) \sim \omega^{1/2}$. We analytically derive scaling laws for the dispersion and vibrational density of states near yielding, including prefactors. These findings reveal that structural order gives rise to a distinct mechanism of mechanical instability---directionally extended multimode softening---contrasting with the localized soft-mode mechanism found in amorphous solids.

{\it Theory---}
The system consists of $N_x \times N_y$ particles with diameter $d$ arranged on a two-dimensional triangular lattice with periodic boundaries, as shown in Fig.~\ref{figmain:setup}.
The initial overlap between neighboring particles is $hd$, which results in the system sizes $L_x=(1-h)dN_x$ and $L_y=\sqrt{3}(1-h)dN_y/2$.
A quasistatic shear strain $\gamma$ is applied.
Let $\bm{s}_1=(1-h)d(1,0)$ and 
$\bm{s}_2=(1-h)d\left(1/2+\gamma\sqrt{3}/2,\sqrt{3}/2\right)$ be the sheared primitive lattice vectors.
The reference position of particle $p$ is $\bm{R}_{p}=n_1\bm{s}_1+n_2\bm{s}_{2}$ ($n_1\in \{1,\ldots, N_x\}, n_2\in \{1,\ldots, N_y\}$), and $\bm{r}_{p}$ denotes the dynamic position.
Each particle has mass $m$, and the interparticle force between particles $p$ and $q$ is given by the Hertzian contact model $\bm{f}_{pq}=-\kappa (d-r_{pq})^{3/2}\bm{r}_{pq}\Theta(d-r_{pq})/r_{pq}$, where $\bm{r}_{pq}$ is the relative position vector, $r_{pq}=|\bm{r}_{pq}|$, $\kappa$ is the contact stiffness, and $\Theta$ is the Heaviside step function.
The equation of motion for particle $p$ is
$
m \ddot{r}_{p,a}
= \sum_{q \neq p} f_{pq,a}$.
We label a representative particle as $0$ and denote its six nearest neighbors by $q'=1,\dots,6$ (Fig.~\ref{figmain:setup}).
The reference relative positions of the six nearest neighbors are
\begin{equation}
(R_{0q',x},R_{0q',y})
=(1-h)d\left(
\cos\phi_{0q'} + \gamma\sin\phi_{0q'},
\sin\phi_{0q'}
\right),
\label{eqmain:positions}
\end{equation}
with $\phi_{0q'}=(q'-1)\pi/3$.
The shear stress at the reference configuration specified by $h$ and $\gamma$ is obtained from the virial formula; exploiting lattice symmetry, we reduce it to a sum over three symmetry-independent nearest-neighbor pairs,
\begin{equation}
\sigma_{ab}
= -\frac{2}{\sqrt{3}(1-h)^2 d^3}
\sum_{q'=1}^{3} R_{0q',a} f_{0q',b},
\label{eqmain:stress}
\end{equation}
where $a, b\in\{x, y\}$ and $f_{0q',a}=-\kappa (d-R_{0q'})^{3/2}R_{0q',a}\Theta(d-R_{0q'})/R_{0q'}$ with $R_{0q'}=|\bm{R}_{0q'}|$, and out-of-plane depth $d$ is assumed for dimensional consistency.

\begin{figure}
    \centering
    \includegraphics[width=0.9\linewidth]{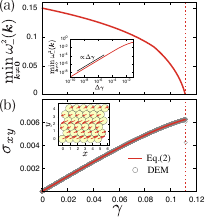}
    \caption{
    (a) Minimum Hessian eigenvalue $\min_{\bm{k}\ne \bm{0}}\omega_{-}^{2}(\bm{k})$ [Eq.~\eqref{eqmain:eigenvalue}], evaluated numerically, as a function of shear strain $\gamma$ for a system with $N_x=N_y=6$ and $h=0.05$. 
    The minimum eigenvalue decreases monotonically and defines a yield strain, $\gamma_c$, where it vanishes (red dotted line, also shown in (b)).
    Inset: Logarithmic plot of $\omega_{-}^{2}$ versus the distance to yielding, $\Delta\gamma=\gamma_c-\gamma$, demonstrating the linear scaling $\omega_{-}^{2}\sim\Delta\gamma$ near yielding.
    (b) Shear stress $\sigma_{xy}$ [Eq.~\eqref{eqmain:stress}], evaluated numerically (red curve), as a function of $\gamma$ for the same system, in quantitative agreement with DEM simulation results (gray symbols). The stress drop in the DEM data occurs at the numerically predicted $\gamma_c$ (dotted line), validating the numerical framework. Inset: Numerical eigenmode [Eq.~\eqref{eqmain:eigenvector}] associated with the minimum eigenvalue at yielding, showing a plane-wave character (red arrows).}
    \label{figmain:rheology}
\end{figure}

\begin{figure*}
    \centering
    \includegraphics[width=1.0\linewidth]{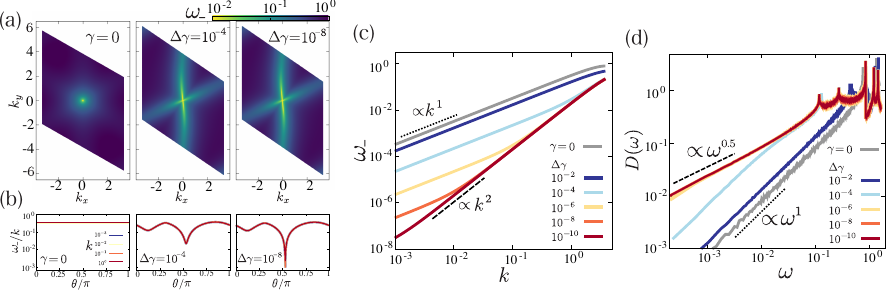}
    \caption{
    (a) Two-dimensional dispersion relations $\omega_{-}(k_x,k_y)$ at $\gamma=0$ and near yielding ($\Delta\gamma=10^{-4},10^{-8}$). Color indicates $\omega_{-}$. At $\gamma=0$, low-frequency modes are concentrated near $k=0$. Near yielding, a cross-shaped low-frequency region appears, indicating directional soft modes.
    (b) Angular dependence of $\omega_{-}/k$ at several $k$ for various $\Delta\gamma$. Near yielding ($\Delta\gamma\to 0$), a minimum develops around $\theta/\pi\simeq0.5$. The critical angle $\theta_c$ is defined as the angle at which the minimum occurs at yielding.
    (c) Dispersion relation along the direction $\theta_c$, $\omega_{-}$ vs $k$, for various $\Delta\gamma$. Far from yielding, $\omega_{-}\sim k$; near yielding, the dispersion crosses over to $\omega_{-}\sim k^2$. (d) Vibrational density of states for various $\Delta\gamma$, computed from all $\omega_{\pm}(\bm{k})$. Far from yielding, $D(\omega)\sim\omega$ (Debye scaling); near yielding, $D(\omega)\sim\omega^{1/2}$.
    Panels (a)--(c) are obtained numerically from Eq.~\eqref{eqmain:eigenvalue} in the continuum limit ($N_{x}, N_{y}\to\infty$), while (d) uses a large system $N_{x}=N_{y}=4000$.
    The scalings $\omega_{-}\sim k^2$ and $D(\omega)\sim\omega^{1/2}$ are analytically derived, including prefactors, in the Appendix (Eqs.~\eqref{eqmain:scaled_omega} and \eqref{eqmain:analytical_vdos}).
    }
    \label{figmain:dispersion}
\end{figure*}

\begin{figure}
    \centering
    \includegraphics[width=0.9\linewidth]{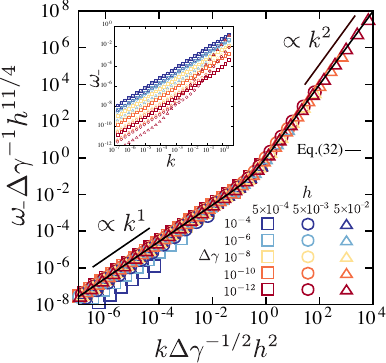}
    \caption{
    Scaled dispersion relations near yielding for various $\Delta\gamma$ and $h$, plotted as $\omega_{-}\Delta\gamma^{-1}h^{11/4}$ versus $k\Delta\gamma^{-1/2}h^{2}$, showing data collapse. Symbols denote numerical results obtained from Eq.~\eqref{eqmain:eigenvalue}.
    The data exhibit linear ($\omega_{-}\sim k$) and quadratic ($\omega_{-}\sim k^2$) regimes with a crossover at $k\Delta\gamma^{-1/2}h^{2}\sim 1$. The black curve shows the analytical result [Eq.~\eqref{eqmain:scaled_omega}], in quantitative agreement with the numerical data. Inset: Unscaled dispersion relations for different $\Delta\gamma$ and $h$ before rescaling.
    }
    \label{figmain:dispersion-scaled}
\end{figure}

To compute the vibrational properties of the system, we analyze the eigenvalues and eigenvectors of the Hessian matrix.
Linearizing the equations of motion around the reference configuration in terms of small displacements $u_{p,a}$ yields
\begin{equation}
m\ddot{u}_{p,a}
= -\sum_{q\ne p} K_{pq,ab}(u_{p,b}-u_{q,b}),
\label{eqmain:eom_perturbation}
\end{equation}
where the matrix $K_{pq,ab}$ corresponds to the Hessian matrix element and is given by
\begin{equation}
K_{pq,ab}
=\kappa\left[
\frac{3}{2}\xi_{pq}^{1/2} P_{\parallel,ab}
-\frac{\xi_{pq}^{3/2}}{R_{pq}} P_{\perp,ab}
\right]\Theta(\xi_{pq}),
\label{eqmain:Kpq}
\end{equation}
with $R_{pq}=|\bm{R}_{pq}|$, $R_{pq,a}=R_{q,a}-R_{p,a}$, $\xi_{pq}=d-R_{pq}$,  $P_{\parallel,ab}=\nu_{pq,a}\nu_{pq,b}$, $P_{\perp,ab}=\delta_{ab}-\nu_{pq,a}\nu_{pq,b}$, $\nu_{pq,a}=R_{pq,a}/R_{pq}$, and $\delta_{ab}$ the Kronecker delta.
Equations~\eqref{eqmain:eom_perturbation} and \eqref{eqmain:Kpq} are general for Hertzian particle systems; in the present crystal, they can be further simplified by symmetry.
Let us focus on a representative particle ($p=0$).
Using Eq.~\eqref{eqmain:positions}, Eq.~\eqref{eqmain:eom_perturbation} reduces to
\begin{equation}
m\ddot{u}_{0,a}
= -\sum_{q'=1}^{6} K_{0q',ab}(u_{0,b}-u_{q',b}).
\end{equation}
We consider plane-wave perturbations of the form
\begin{equation}
u_{p,a}=U_a\,e^{i(\bm{k}\cdot \bm{R}_{p}-\omega t)},
\label{eqmain:plane_wave}
\end{equation}
where the wave vector $\bm{k}$ is quantized by the periodic boundary conditions via reciprocal lattice vectors $\bm{b}_{i}$ defined by $\bm{b}_{i}\cdot\bm{s}_j=2\pi\delta_{ij}$.
Substituting Eq.~\eqref{eqmain:plane_wave} into Eq.~\eqref{eqmain:eom_perturbation} and grouping pairs of opposite neighbors, $(1, 4)$, $(2, 5)$, and $(3, 6)$, we obtain
\begin{equation}
\omega^2 U_a
=\frac{2}{m}\sum_{q'=1}^{3}
\bigl[1-\cos (\bm{k}\cdot \bm{R}_{0q'})\bigr]
K_{0q',ab}U_b
= \Lambda_{ab}U_b,
\label{eqmain:eigenvalue_equation}
\end{equation}
where
$\Lambda_{ab}\equiv (2/m)\sum_{q'=1}^3 \bigl[1-\cos (\bm{k}\cdot \bm{R}_{0q'})\bigr]\,K_{0q',ab}$.
For a given wave vector $\bm{k}$, the eigenvalues of the matrix $\Lambda$ with components $\Lambda_{ab}$ are
\begin{equation}
\omega^2_{\pm}
=\frac{\tr\Lambda\pm\sqrt{(\tr\Lambda)^2-4\det\Lambda}}{2},
\label{eqmain:eigenvalue}
\end{equation}
with $\tr\Lambda=\Lambda_{xx}+\Lambda_{yy}$, and $\det\Lambda=\Lambda_{xx}\Lambda_{yy}-\Lambda_{xy}\Lambda_{yx}$.
Here, $\Lambda_{ab}$ depends on $h$ and $\gamma$, which specify the system, together with the wave vector magnitude $k$ and polar angle $\theta$, and thus $\omega^2_{\pm}=\omega^2_{\pm}(k, \theta, \gamma; h)$, where the allowed values of $(k, \theta)$ depend on the system size.
The corresponding dimensionless eigenvectors are
\begin{equation}
(U_{\pm, x}, U_{\pm, y})
=\frac{1}{\sqrt{\Lambda_{xy}^2+(\Lambda_{xx}-\omega^2_{\pm})^2}}
\left(-\Lambda_{xy}, \Lambda_{xx}-\omega^2_{\pm}\right).
\label{eqmain:eigenvector}
\end{equation}
The $\bm{k}=\bm{0}$ mode corresponding to uniform translation is excluded throughout.
The results shown in Figs.~\ref{figmain:rheology}, \ref{figmain:dispersion}, and \ref{figmain:dispersion-scaled} are obtained numerically from Eqs.~\eqref{eqmain:stress}, \eqref{eqmain:eigenvalue}, and \eqref{eqmain:eigenvector}. We choose units $d=m=\kappa=1$ in all figures.
Throughout this Letter, the terms numerical, analytical, and DEM simulation refer to evaluations of the exact model equations [e.g., Eqs.~\eqref{eqmain:stress} and \eqref{eqmain:eigenvalue}], asymptotic expressions derived in the Appendix, and discrete element method simulation (see Appendix for details), respectively.

Before turning to the full eigenvalue spectrum near yielding, we consider the minimum Hessian eigenvalue and its associated eigenvector—quantities extensively studied in amorphous systems~\cite{Maloney2004-wh, Maloney2006-xw}—as well as the shear stress.
The minimum eigenvalue obtained numerically from Eq.~\eqref{eqmain:eigenvalue} for $N_x = N_y = 6$ and $h = 0.05$ is shown as a function of the shear strain $\gamma$ in Fig.~\ref{figmain:rheology}(a).
The minimum eigenvalue decreases monotonically and defines a yield strain $\gamma_c$ where it vanishes (red dotted line). In the continuum limit, the yield strain is analytically obtained [Eq.~\eqref{eqmain:gammac_expansion}, see Appendix], except for minor system size effects.
The inset shows the minimum $\omega_{-}^2$ as a function of the distance to yielding, $\Delta\gamma=\gamma_c-\gamma$, revealing a linear scaling $\omega_{-}^2\sim\Delta\gamma$. This scaling is confirmed analytically in the Appendix [see Eq.~\eqref{eqmain:lambda_expansion_yielding}].
This scaling contrasts with the behavior in amorphous solids, where $\omega_{-}^2 \sim \Delta\gamma^{1/2}$ due to non-affine deformation~\cite{Maloney2004-wh, Maloney2006-xw}. In the present crystalline system, such non-affine effects are absent by symmetry~\cite{Maloney2006-xw}.
Figure~\ref{figmain:rheology}(b) shows the numerical $\sigma_{xy}$ [Eq.~\eqref{eqmain:stress}] as a function of $\gamma$ (red curve), alongside DEM simulation results (gray symbols). The numerical results reproduce the stress–strain response up to yielding, and the numerically predicted $\gamma_c$ (dotted line) coincides with the strain at which the stress drop occurs in the DEM data.
The inset shows the eigenvector corresponding to the minimum eigenvalue at $\gamma_c$, represented by red arrows. This mode exhibits a plane-wave character and is qualitatively different from the spatially localized instability modes observed in amorphous solids~\cite{Maloney2004-wh, Maloney2006-xw}, reflecting the crystalline symmetry of the system.

We now turn to the key question of how the full vibrational spectrum evolves as yielding is approached. Figure~\ref{figmain:dispersion}(a) shows the numerical two-dimensional dispersion relation $\omega_{-}(k_x,k_y)$, with color indicating $\omega_{-}$, in the continuum limit ($N_{x}, N_{y}\to\infty$) for $h=0.05$ at zero shear ($\gamma=0$) and near yielding ($\Delta\gamma=10^{-4}$ and $10^{-8}$).
At $\gamma=0$, low-frequency modes are confined near the origin in wave number space, consistent with a linear (acoustic) dispersion relation $\omega_{-} \sim k$. As yielding is approached ($\Delta\gamma \to 0$), a distinct cross-shaped low-frequency regime emerges, indicating the development of multiple soft modes along specific wave number directions. This multimode softening contrasts with the single-mode localized instability found in amorphous solids~\cite{Maloney2004-wh, Maloney2006-xw} and has not, to our knowledge, been reported in athermal solids. To characterize the directionality of this softening, Fig.~\ref{figmain:dispersion}(b) shows the numerical $\omega_{-}/k$ as a function of the polar angle $\theta/\pi$ for several values of $k=|\bm{k}|$. At $\gamma=0$, $\omega_{-}/k$ shows only weak angular dependence. Near yielding, a pronounced minimum develops around $\theta/\pi\simeq0.5$, demonstrating that soft modes concentrate along this direction. We define $\theta_{c}$ as the angle at which $\omega_{-}$ attains its minimum at yielding, which is obtained analytically (see Eq.~\eqref{eqmain:thetac_expansion}).

Along the direction $\theta_c$, the dispersion relation changes dramatically near yielding. Figure~\ref{figmain:dispersion}(c) shows the numerical $\omega_{-}$ as a function of $k$ in the $\theta_c$ direction, for the unsheared state ($\gamma=0$) and for several values of $\Delta\gamma$ at $h=0.05$. Far from yielding, the dispersion is linear, $\omega_{-} \sim k$. As yielding is approached, the linear contribution is progressively suppressed and a quadratic dispersion, $\omega_{-} \sim k^2$, becomes dominant, indicating that long-wavelength acoustic waves become anomalously dispersive—propagating at different speeds depending on wavelength, unlike conventional acoustic waves. To characterize the resulting low-frequency excitations, we analyze the vibrational density of states (VDOS), $D(\omega)=(2N_xN_y)^{-1}\sum_i \delta(\omega-\omega_i)$~\cite{Mizuno2017-gh, Petit2025-vz}, with $\omega_{i}$ ($i=1,\ldots,2N_{x}N_{y}$) being the eigenfrequency of each vibrational mode. Using the full eigenvalue spectrum from a large system with $N_x=N_y=4000$, we compute the VDOS for $\gamma=0$ and various $\Delta\gamma$, as shown in Fig.~\ref{figmain:dispersion}(d).
Far from yielding, the Debye scaling $D(\omega)\sim\omega$ holds, whereas close to yielding, a non-Debye scaling $D(\omega)\sim\omega^{1/2}$ emerges, indicating a large number of low-frequency soft modes. These anomalous scaling laws are analytically derived, including the prefactors, in the Appendix (Eqs.~\eqref{eqmain:scaled_omega} and \eqref{eqmain:analytical_vdos}).

Near yielding, the anomalous feature in the dispersion relation can be fully characterized using $h$ and $\Delta\gamma$.
The numerical dispersion relations in the direction of the critical angle $\theta_{c}$ for various $\Delta\gamma$ and $h$ are shown as symbols in Fig.~\ref{figmain:dispersion-scaled}.
Near yielding, the data collapse when plotted as $\omega_{-}\,\Delta\gamma^{-1}h^{11/4}$ versus $k\,\Delta\gamma^{-1/2}h^{2}$ (in units $m=d=\kappa=1$); the unscaled data 
are shown in the inset for comparison.
The analytical result (Eq.~\eqref{eqmain:scaled_omega}) quantitatively reproduces the numerical data, including the non-trivial scaling exponents and prefactors; minor deviations are attributable to higher-order corrections in $\Delta\gamma$ and $h$. The crossover from $\omega_{-}\sim k$ to $\omega_{-}\sim k^2$ occurs at $k \sim \Delta\gamma^{1/2}/h^{2}$, indicating that the quadratic dispersion $\omega_{-}\sim k^2$ persists down to arbitrarily long wavelengths as $\Delta\gamma\to 0$. The crossover wave number defines a characteristic length $l_c=h^2\Delta\gamma^{-1/2}$, which diverges as $\Delta\gamma\to 0$, signaling a growing length scale near yielding.

Before the conclusion, let us discuss the generality of our findings. For a broad class of interparticle potentials, dimensions, and crystal structures, a similar framework can be applied, leading to an expansion near yielding of the form $\omega_{-}^2=A\Delta\gamma k^2+Bk^4$ (see Eq.~\eqref{eqmain:lambda_expansion_yielding}). This implies that the anomalous scaling $\omega_{-}\sim k^2$ near yielding, and the corresponding non-Debye vibrational density of states, are generic features, with potential- and structure-dependent prefactors (we note that 
the continuous softening $\omega_{-}^2\sim\Delta\gamma$ requires nonlinear 
interparticle potentials; e.g., Hookean interactions would exhibit no such 
softening). As an example, both the $\omega_{-}\sim k^2$ scaling and the data collapse of $\omega_{-}\Delta\gamma^{-1}$ versus $k\Delta\gamma^{-1/2}$ are confirmed for the Weeks--Chandler--Andersen potential (Fig.~\ref{figmain:dispersion_wca}).

{\it Conclusion---}
In summary, we have investigated shear yielding in an athermal crystal by analyzing the Hessian eigenvalue spectrum in wave number space. We have shown that the present system exhibits directionally extended multimode softening in the direction of the critical angle $\theta_c$. Near yielding, the conventional acoustic dispersion $\omega\sim k$ is replaced by $\omega\sim k^2$, the Debye scaling $D(\omega)\sim \omega$ is replaced by $D(\omega)\sim \omega^{1/2}$, and a diverging length scale $l_c$ emerges. Importantly, we analytically derive these scaling laws, including prefactors. These findings reveal a mechanism of mechanical instability in ordered athermal solids that is fundamentally distinct from the localized soft-mode scenario in amorphous systems~\cite{Maloney2004-wh, Maloney2006-xw, Xu2017-dt}, suggesting that structural order plays a crucial role in shaping the nature of yielding. These results provide a foundation for understanding and controlling the mechanical failure of ordered particulate systems, from colloidal crystals to granular packings.

\section*{Acknowledgments}
This work was supported by JSPS KAKENHI Grant Numbers JP24KJ0156, JP25K17359, JP24H00196, JP23K03248, and JP22K03459, and JST ERATO Grant Number JPMJER2401.
The computation in this work has been done using the facilities of the Supercomputer Center, the Institute for Solid State Physics, the University of Tokyo (2025-Ba-0063).

\bibliographystyle{apsrev4-2}
\bibliography{ref}

@BOOK{Andreotti2013-yo,
  title     = "Granular Media: Between Fluid and Solid",
  author    = "Andreotti, Bruno and Forterre, Yo{\"e}l and Pouliquen, Olivier",
  publisher = "Cambridge University Press",
  month     =  jun,
  year      =  2013,
}

@ARTICLE{Daniels2005-wn,
  title     = "Hysteresis and competition between disorder and crystallization
               in sheared and vibrated granular flow",
  author    = "Daniels, Karen E and Behringer, Robert P",
  journal   = "Phys. Rev. Lett.",
  publisher = "American Physical Society (APS)",
  volume    =  94,
  number    =  16,
  pages     =  168001,
  month     =  apr,
  year      =  2005,
  language  = "en"
}

@ARTICLE{Pichard2016-ad,
  title     = "Surface waves in granular phononic crystals",
  author    = "Pichard, H and Duclos, A and Groby, J-P and Tournat, V and Zheng,
               L and Gusev, V E",
  journal   = "Phys. Rev. E",
  publisher = "APS",
  volume    =  93,
  number    =  2,
  pages     =  023008,
  month     =  feb,
  year      =  2016,
  language  = "en"
}

@ARTICLE{Kool2022-qp,
  title     = "Gardner-like crossover from variable to persistent force contacts
               in granular crystals",
  author    = "Kool, Lars and Charbonneau, Patrick and Daniels, Karen E",
  journal   = "Phys. Rev. E",
  publisher = "American Physical Society",
  volume    =  106,
  number    =  5,
  pages     =  054901,
  month     =  nov,
  year      =  2022,
  language  = "en"
}

@BOOK{Kittel2004-zm,
  title     = "Introduction to solid state physics: International edition",
  author    = "Kittel, Charles",
  publisher = "John Wiley \& Sons",
  edition   =  8,
  month     =  dec,
  year      =  2004,
}

@ARTICLE{Maloney2004-wh,
  title     = "Universal breakdown of elasticity at the onset of material
               failure",
  author    = "Maloney, Craig and Lemaître, Anaël",
  journal   = "Phys. Rev. Lett.",
  publisher = "American Physical Society (APS)",
  volume    =  93,
  number    =  19,
  pages     =  195501,
  month     =  nov,
  year      =  2004,
  language  = "en"
}

@ARTICLE{Petit2025-vz,
  title     = "Vibrational similarities in jamming-unjamming of polycrystalline
               and disordered granular packings",
  author    = "Petit, Juan C and Ganguly, Saswati and Sperl, Matthias",
  journal   = "Phys. Rev. Res.",
  publisher = "American Physical Society (APS)",
  volume    =  7,
  number    =  3,
  pages     =  033040,
  month     =  jul,
  year      =  2025,
  language  = "en"
}

@ARTICLE{Maloney2006-xw,
  title    = "Amorphous systems in athermal, quasistatic shear",
  author   = "Maloney, Craig E and Lemaître, Anaël",
  journal  = "Phys. Rev. E",
  volume   =  74,
  pages    =  016118,
  month    =  jul,
  year     =  2006,
  language = "en"
}

@ARTICLE{Nakai2025-lk,
  title     = "Dislocation glides in granular media",
  author    = "Nakai, Fumiaki and Uneyama, Takashi and Sasaki, Yuto and Yoshii,
               Kiwamu and Katsuragi, Hiroaki",
  journal   = "Phys. Rev. Lett.",
  publisher = "American Physical Society (APS)",
  volume    =  135,
  number    =  4,
  pages     =  048202,
  month     =  jul,
  year      =  2025,
  language  = "en"
}

@ARTICLE{Nicolas2018-zj,
  title     = "Deformation and flow of amorphous solids: Insights from
               elastoplastic models",
  author    = "Nicolas, Alexandre and Ferrero, Ezequiel E and Martens, Kirsten
               and Barrat, Jean-Louis",
  journal   = "Rev. Mod. Phys.",
  publisher = "American Physical Society (APS)",
  volume    =  90,
  number    =  4,
  pages     =  045006,
  month     =  dec,
  year      =  2018,
  language  = "en"
}

@ARTICLE{Mouhat2014-cx,
  title     = "Necessary and sufficient elastic stability conditions in various
               crystal systems",
  author    = "Mouhat, Félix and Coudert, François-Xavier",
  journal   = "Phys. Rev. B",
  publisher = "American Physical Society",
  volume    =  90,
  number    =  22,
  pages     =  224104,
  month     =  dec,
  year      =  2014,
  language  = "en"
}

@ARTICLE{Tanguy2010-fi,
  title   = "Vibrational modes as a predictor for plasticity in a model glass",
  author  = "Tanguy, A and Mantisi, B and Tsamados, M",
  journal = "Europhys. Lett.",
  volume  =  90,
  number  =  1,
  pages   =  16004,
  month   =  apr,
  year    =  2010
}

@ARTICLE{Xu2017-dt,
  title     = "Instabilities of jammed packings of frictionless spheres under
               load",
  author    = "Xu, Ning and Liu, Andrea J and Nagel, Sidney R",
  journal   = "Phys. Rev. Lett.",
  publisher = "American Physical Society",
  volume    =  119,
  number    =  21,
  pages     =  215502,
  month     =  nov,
  year      =  2017,
  language  = "en"
}

@ARTICLE{Mizuno2017-gh,
  title     = "Continuum limit of the vibrational properties of amorphous solids",
  author    = "Mizuno, Hideyuki and Shiba, Hayato and Ikeda, Atsushi",
  journal   = "Proc. Natl. Acad. Sci. U. S. A.",
  publisher = "National Academy of Sciences",
  volume    =  114,
  number    =  46,
  pages     = "E9767--E9774",
  month     =  nov,
  year      =  2017,
  language  = "en"
}

@ARTICLE{Otsuki2023-vx,
  title    = "An exact expression of three-body system for the complex shear
              modulus of frictional granular materials",
  author   = "Otsuki, Michio and Hayakawa, Hisao",
  journal  = "Soft Matter",
  volume   =  19,
  number   =  11,
  pages    = "2127--2137",
  month    =  mar,
  year     =  2023,
  language = "en"
}

@ARTICLE{Bragg1947-ua,
  title     = "A dynamical model of a crystal structure",
  author    = "Bragg, L and Nye, J F",
  journal   = "Proc. R. Soc. Lond.",
  publisher = "The Royal Society",
  volume    =  190,
  number    =  1023,
  pages     = "474--481",
  month     =  sep,
  year      =  1947,
  language  = "en"
}

@ARTICLE{Bragg1949-km,
  title     = "A dynamical model of a crystal structure. {II}",
  author    = "Bragg, L and Lomer, W M",
  journal   = "Proc. R. Soc. Lond.",
  publisher = "The Royal Society",
  volume    =  196,
  number    =  1045,
  pages     = "171--181",
  month     =  mar,
  year      =  1949,
  language  = "en"
}

@ARTICLE{Krishnan2022-oi,
  title     = "Universal non-Debye low-frequency vibrations in sheared amorphous solids",
  author    = "Krishnan, Vishnu V. and Ramola, Kabir and Karmakar, Smarajit",
  journal   = "Soft Matter",
  publisher = "Royal Society of Chemistry (RSC)",
  volume    =  18,
  number    =  17,
  pages     = "3395--3402",
  month     =  may,
  year      =  2022,
  language  = "en"
}

@ARTICLE{Merkel2010-kz,
  title     = "Dispersion of elastic waves in three-dimensional noncohesive
               granular phononic crystals: properties of rotational modes",
  author    = "Merkel, A and Tournat, V and Gusev, V",
  journal   = "Phys. Rev. E",
  publisher = "American Physical Society (APS)",
  volume    =  82,
  pages     =  031305,
  month     =  sep,
  year      =  2010,
  language  = "en"
}

@ARTICLE{Merkel2011-xv,
  title     = "Experimental evidence of rotational elastic waves in granular
               phononic crystals",
  author    = "Merkel, A and Tournat, V and Gusev, V",
  journal   = "Phys. Rev. Lett.",
  publisher = "American Physical Society (APS)",
  volume    =  107,
  number    =  22,
  pages     =  225502,
  month     =  nov,
  year      =  2011,
  language  = "en"
}

@ARTICLE{Karuriya2023-me,
  title     = "Granular crystals as strong and fully dense architectured
               materials",
  author    = "Karuriya, Ashta Navdeep and Barthelat, Francois",
  journal   = "Proc. Natl. Acad. Sci. U. S. A.",
  publisher = "National Academy of Sciences",
  volume    =  120,
  number    =  1,
  pages     = "e2215508120",
  month     =  jan,
  year      =  2023,
  language  = "en"
}

@ARTICLE{Karuriya2024-cb,
  title     = "Plastic deformations and strain hardening in fully dense granular
               crystals",
  author    = "Karuriya, Ashta Navdeep and Barthelat, Francois",
  journal   = "J. Mech. Phys. Solids",
  publisher = "Elsevier BV",
  volume    =  186,
  number    =  105597,
  pages     =  105597,
  month     =  may,
  year      =  2024,
  language  = "en"
}

\appendix
\section*{End Matter}
{\it DEM simulation---}
We use DEM simulations to validate Eqs.~\eqref{eqmain:stress} and \eqref{eqmain:eigenvalue} by comparing the stress–strain response and the yield strain [see Fig.~\ref{figmain:rheology}]. The simulations are performed using LAMMPS with a frictionless Hertzian contact model and Tsuji-type damping. The equation of motion is
\begin{equation}
\begin{split}
m \ddot{r}_{p,a}
= -\sum_{q \neq p} \Bigl[&
\kappa (d-r_{pq})^{3/2} \frac{r_{pq,a}}{r_{pq}}\\
&+ \eta v_{pq,b} \frac{r_{pq,a}r_{pq,b}}{r^2_{pq}}
\Bigr]\Theta(d-r_{pq}),
\end{split}
\label{eqmain:eom_dem}
\end{equation}
where $v_{pq,a}=v_{q,a}-v_{p,a}$. The damping coefficient is given by $\eta=\alpha \sqrt{m \kappa (d-r_{pq})^{1/2}/2}$ with
$\alpha
=1.2728
- 4.2783 e
+ 11.087 e^2
- 22.348 e^3
+ 27.467 e^4
- 18.022 e^5
+ 4.8218 e^6$, where the restitution coefficient is set to $e=0.5$, giving $\alpha\simeq 0.34074$. In the small strain-rate regime considered here, the resulting stress is insensitive to this choice. Following the main text, we choose units such that $d=m=\kappa=1$. The system consists of $N_x=N_y=6$ particles initially arranged on a triangular lattice with an overlap parameter $h=0.05$ under periodic boundary conditions. Small random displacements of amplitude $10^{-8}d$ are added to the initial particle positions as a perturbation. Simple shear is imposed by tilting the simulation box using the {\tt deform} command. The shear rate is set to $\dot{\gamma}=10^{-8}$ to ensure a quasistatic regime, and the time step size is fixed to $0.1$ in dimensionless units. The shear stress is computed as
\begin{equation}
\sigma_{ab}
= -\frac{1}{L_xL_yd}\left[
\sum_{p} m \dot{r}_{p,a}\dot{r}_{p,b}
+ \sum_{p\neq q} \frac{1}{2} r_{pq,a} f_{pq,b}
\right],
\end{equation}
where the kinetic contribution is negligible in the present low-$\dot{\gamma}$ regime.

{\it Yield strain $\gamma_c$ and critical angle $\theta_c$---}
Equation~\eqref{eqmain:eigenvalue} provides the full eigenvalue spectrum on a discretized $k$ grid.
In the vicinity of yielding,
where $\omega_{-}^2$ becomes small, this expression can be reduced within a continuum approximation for $k$. Here we consider only $\omega_{-}^2$. Near yielding, $\det\Lambda$ becomes small and $\omega_{-}^2$ can be reduced to $\omega_{-}^2=\det\Lambda / \tr \Lambda$. Writing the wave vector in polar coordinates, $k_x=k\cos\theta$ and $k_y=k\sin\theta$, and expanding Eq.~\eqref{eqmain:eigenvalue_equation} to $k^4$ order, we obtain
\begin{equation}
\begin{split}
    \Lambda_{ab}
    &\simeq \frac{2}{m}
    \sum_{q'=1}^3
    \left(\frac{\mu_{q'}^2}{2}k^2
    -\frac{\mu_{q'}^4}{24}k^4
    \right)
    \,K_{0q',ab}\\
    &=\Lambda_{2,ab}(\gamma, \theta; h)k^2-\Lambda_{4,ab}(\gamma, \theta; h)k^4
\end{split}
\end{equation}
where $\mu_{q'}=R_{0q',x}\cos\theta + R_{0q',y}\sin\theta$ and 
$\theta\in(0, \pi)$.
Using this expansion, the approximate eigenvalue $\omega_{-}^2$ can be expanded as follows:
\begin{equation}
    \omega_{-}^2=\Omega_{2}(\theta, \gamma; h)k^2
    +\Omega_{4}(\theta, \gamma; h)k^4
    \label{eqmain:expansion_k}
\end{equation}
with
\begin{align}
    \Omega_{2}&=\frac{\det\Lambda_{2}}{\tr\Lambda_{2}}
    \label{eqmain:Omega2}\\
    \Omega_{4}&=-\frac{[(\tr\Lambda_{2})^2
    -\det\Lambda_{2}]\tr\Lambda_{4}
    -\tr\Lambda_{2}\tr(\Lambda_{2}\Lambda_{4})}{(\tr\Lambda_{2})^2}
    \label{eqmain:Omega4}
\end{align}
We assume that yielding is governed by a long-wavelength instability,
identified by the vanishing of the $k^2$ coefficient $\Omega_2$
in Eq.~\eqref{eqmain:expansion_k}, and thus focus on $\Omega_{2}(\theta, \gamma; h)$.
At the yielding strain, the minimum eigenvalue vanishes along the direction $\theta_c$, leading to $\Omega_{2}(\theta_c,\gamma_c)=0$ and
$\partial_{\theta}\Omega_{2}(\theta_c,\gamma_c)=0$.
Using $\Omega_2=\det\Lambda_2/\tr\Lambda_2$, these conditions reduce to
\begin{align}
    \det\Lambda_{2}\big|_{\theta=\theta_c,\gamma
    =\gamma_c}&=0,\label{eqmain:lambda_0}\\
    \partial_{\theta}\det\Lambda_{2}\big|_{\theta
    =\theta_c,\gamma=\gamma_c}&=0.
    \label{eqmain:lambda_theta_0}
\end{align}
Using the identity, $(a\cos\theta+b\sin\theta)^2=(a^2+b^2)/2+(a^2-b^2)\cos 2\theta/2+ab\sin 2\theta$, to separate the $\theta$ dependence, $\Lambda_{2,ab}$ can be expressed as
\begin{equation}
    \Lambda_{2,ab}(\gamma, \theta; h)=A_{ab}+B_{ab}\cos2\theta+C_{ab}\sin2\theta
\end{equation}
with
\begin{align}
    A_{ab}(\gamma; h)&=\sum_{q'=1}^{3}
    \frac{R_{0q',x}^2+R_{0q',y}^2}{4}K_{0q',ab},\\
    B_{ab}(\gamma; h)&=\sum_{q'=1}^{3}
    \frac{R_{0q',x}^2-R_{0q',y}^2}{4}K_{0q',ab},\\
    C_{ab}(\gamma; h)&=\sum_{q'=1}^{3}
    \frac{R_{0q',x}R_{0q',y}}{2}K_{0q',ab}.
\end{align}
Using this relation, we obtain
\begin{equation}
\begin{split}
    \det\Lambda_{2}
    =&\alpha_{0}+\alpha_{2c}\cos2\theta
    +\alpha_{2s}\sin2\theta\\
    &+\alpha_{4c}\cos4\theta
    +\alpha_{4s}\sin4\theta
\end{split}
\end{equation}
with $\alpha_{0}=\det A + (\det B+\det C)/2$, $\alpha_{2c}=(\tr A)(\tr B)-(\tr AB)$, $\alpha_{2s}=(\tr A)(\tr C)-(\tr AC)$, $\alpha_{4c}=(\det B-\det C)/2$, and $\alpha_{4s}=[(\tr B)(\tr C)-(\tr BC)]/2$.
We also obtain
\begin{equation}
\begin{split}
    \partial_{\theta}(\det\Lambda_{2})
    =&-2\alpha_{2c}\sin2\theta
    +2\alpha_{2s}\cos2\theta\\
    &-4\alpha_{4c}\sin4\theta
    +4\alpha_{4s}\cos4\theta.
\end{split}
\end{equation}
Introducing $\chi=\tan\theta$, we obtain
\begin{equation}
\det\Lambda_{2}
=
\frac{
\alpha'_4\chi^4
+\alpha'_3\chi^3
+\alpha'_2\chi^2
+\alpha'_1\chi
+\alpha'_0
}{
(1+\chi^2)^2
},
\label{eqmain:lambda_expansion}
\end{equation}
where
$\alpha'_4=\alpha_0-\alpha_{2c}+\alpha_{4c}$,
$\alpha'_3=2\alpha_{2s}-4\alpha_{4s}$,
$\alpha'_2=2\alpha_0-6\alpha_{4c}$,
$\alpha'_1=2\alpha_{2s}+4\alpha_{4s}$,
and
$\alpha'_0=\alpha_0+\alpha_{2c}+\alpha_{4c}$.
Moreover, we obtain
\begin{equation}
    \partial_{\theta}(\det\Lambda_{2})=
    \frac{\alpha''_4\chi^4+\alpha''_3\chi^3+\alpha''_2\chi^2+\alpha''_1\chi+\alpha''_0}{(1+\chi^2)^2},
\label{eqmain:lambda_theta_expansion}
\end{equation}
where
$\alpha''_4=-2\alpha_{2s}+4\alpha_{4s}$,
$\alpha''_3=-4\alpha_{2c}+16\alpha_{4c}$,
$\alpha''_2=-24\alpha_{4s}$,
$\alpha''_1=-4\alpha_{2c}-16\alpha_{4c}$,
and $\alpha''_0=2\alpha_{2s}+4\alpha_{4s}$.
The conditions in Eqs.~\eqref{eqmain:lambda_0} and \eqref{eqmain:lambda_theta_0}, which determine $\theta_c$ and $\gamma_c$,
reduce to setting the numerators of
Eqs.~\eqref{eqmain:lambda_expansion} and
\eqref{eqmain:lambda_theta_expansion} to zero,
leading to two quartic equations.
Among the four roots, we retain the one that minimizes Eq.~\eqref{eqmain:lambda_expansion}.
Solving these equations yields the continuum-limit values of $\gamma_c$ and $\theta_c$ for a given $h$. These results are consistent with those obtained from the discretized calculation without approximation, apart from minor finite-size effects inherent to the discrete system.

Based on Eqs.~\eqref{eqmain:lambda_expansion} and
\eqref{eqmain:lambda_theta_expansion}, we analytically obtain
the expansions of $\gamma_c$ and $\theta_c$ in terms of $h$ with the aid of Mathematica:
\begin{equation}
\begin{split}
    \gamma_c=&\frac{4}{\sqrt{3}}h
    -\frac{2}{\sqrt{3}}h^2
    -\frac{752\sqrt{3}}{243}h^3\\
    &-\frac{11128\sqrt{3}+65536\sqrt{6}}{2187}h^4
    +\mathcal{O}(h^5),
    \label{eqmain:gammac_expansion}
\end{split}
\end{equation}
and
\begin{align}
    \theta_{c}&=\frac{\pi}{2}
    +\frac{16\sqrt{3}}{9\sqrt{2}}h
    +\frac{128\sqrt{3}-242\sqrt{6}}{81}h^2
    +\mathcal{O}(h^3).
    \label{eqmain:thetac_expansion}
\end{align}
The different truncation orders of $\gamma_c$ and $\theta_c$ follow naturally from requiring Eqs.~\eqref{eqmain:lambda_expansion} and \eqref{eqmain:lambda_theta_expansion} to be satisfied up to $\mathcal{O}(h^3)$.

{\it Expansion near yielding---}
Near the yielding strain $\gamma=\gamma_c-\Delta\gamma$ around the critical angle $\theta=\theta_c+\Delta\theta$, using the conditions $\Omega_2(\theta_c,\gamma_c;h)=0$ and $\partial_{\theta}\Omega_2(\theta_c,\gamma_c;h)=0$, the low-$k$ eigenvalue spectrum $\omega_{-}^2(\gamma, \theta)=\Omega_2(\theta,\gamma;h)k^2
+\Omega_4(\theta,\gamma;h)k^4$ can be expanded as follows:
\begin{equation}
\begin{split}
&\omega_{-}^2(\gamma_c-\Delta\gamma, \theta_c+\Delta\theta; h)\\
\simeq&\left[-\partial_\gamma\Omega_2(\theta_c,\gamma_c;h)\Delta\gamma
+\frac{1}{2}\partial_{\theta}^2\Omega_2(\theta_c,\gamma_c;h)\Delta\theta^2\right]
k^2\\
&+\Omega_4(\theta_c,\gamma_c;h)k^4.
\end{split}
\label{eqmain:lambda_expansion_yielding}
\end{equation}
Equation~\eqref{eqmain:lambda_expansion_yielding} with $\Delta\theta=0$ is consistent with the linear decay of the minimum eigenvalue $\omega_{-}^2\sim\Delta\gamma$, as shown in the inset of Fig.~\ref{figmain:rheology}(a).
Using Eqs.~\eqref{eqmain:lambda_0} and \eqref{eqmain:lambda_theta_0} together with Eqs.~\eqref{eqmain:gammac_expansion} and \eqref{eqmain:thetac_expansion}, 
Eqs.~\eqref{eqmain:Omega2} and \eqref{eqmain:Omega4} lead to
\begin{align}
    \partial_\gamma\Omega_2(\theta_c,\gamma_c;h)
    &=\frac{\partial_{\gamma}\det\Lambda_{2}}{\tr \Lambda_{2}}
    \simeq -\sqrt{\frac{3}{2}}\frac{243}{4096}\frac{\kappa d^{5/2}}{mh^{3/2}},
    \label{eqmain:O2g}\\
    \partial_{\theta}^2\Omega_2(\theta_c,\gamma_c;h)
    &=\frac{\partial_{\theta}^2\det\Lambda_{2}}{\tr \Lambda_{2}}
    \simeq \frac{9}{4}\frac{\kappa d^{5/2}h^{1/2}}{m},\\
    \Omega_4(\theta_c,\gamma_c;h)
    &= \frac{\tr (\Lambda_2\Lambda_4)-\tr\Lambda_{2}\tr\Lambda_{4}}{\tr \Lambda_2}
    \simeq \frac{\kappa d^{9/2}h^{5/2}}{m}.
    \label{eqmain:O4}
\end{align}
Symbolic calculations were performed using Mathematica.

The scaling factors in Fig.~\ref{figmain:dispersion-scaled} follow from Eqs.~\eqref{eqmain:O2g} and \eqref{eqmain:O4}. Along $\theta_c$ ($\Delta\theta=0$), the leading-order dispersion is
\begin{equation}
\begin{split}
    &\omega_{-}(\gamma_c-\Delta\gamma, \theta_c; h)\\
    \simeq&
    \sqrt{\sqrt{\frac{3}{2}}\frac{243}{4096}\frac{\kappa d^{5/2}}{mh^{3/2}}\Delta\gamma k^2
    +\frac{\kappa d^{9/2}h^{5/2}}{m}k^4}\\
    =&\sqrt{\frac{\kappa d^{1/2}}{m}}
    \frac{\Delta\gamma}{h^{11/4}}
    \frac{kd h^2}{\sqrt{\Delta\gamma}}
    \sqrt{\sqrt{\frac{3}{2}}\frac{243}{4096}
    +\left(\frac{kd h^2}{\sqrt{\Delta\gamma}}\right)^2}.
    \label{eqmain:scaled_omega}
\end{split}
\end{equation}
Accordingly, with $m=\kappa=d=1$, the dispersion collapses near yielding when plotted as
$\omega_{-}\Delta\gamma^{-1} h^{11/4}$ versus $k \Delta\gamma^{-1/2}h^2$.

The low-$\omega$ scaling in Fig.~\ref{figmain:dispersion}(d) follows analytically from Eq.~\eqref{eqmain:lambda_expansion_yielding}:
\begin{equation}
    \omega_{-}^2(\gamma_c, \theta_c+\Delta\theta; h)
    \simeq
    \frac{9}{8}\frac{\kappa d^{5/2}h^{1/2}}{m}
    k_{\perp}^2
    +\frac{\kappa d^{9/2}h^{5/2}}{m} k_{\parallel}^4,
    \label{eqmain:omega2_yielding}
\end{equation}
where $k_{\parallel}$ and $k_{\perp}$ are the wave number components parallel and perpendicular to the direction $\theta_c$, respectively.
The area enclosed by $C_1^2 x^2 + C_2^4 y^4 < C_3^2$ is
\begin{equation}
    S=4\int_{0}^{\sqrt{C_{3}}/C_{2}}
    \sqrt{\frac{C_{3}^2-C_{2}^4y^4}{C_{1}^2}}dy
    =\frac{2\sqrt{\pi}\Gamma(1/4)}{3\Gamma(3/4)}
    \frac{C_{3}^{3/2}}{C_{1}C_{2}}.
\end{equation}
Identifying $C_1$ and $C_2$ with the coefficients in Eq.~\eqref{eqmain:omega2_yielding} and setting $C_3=\omega$, the region satisfying $\omega_{-}^2<\omega^2$ has area
\begin{equation}
    S(\omega)\simeq \frac{4\sqrt{2\pi}\Gamma(1/4)}{9\Gamma(3/4)}
    \left(\frac{m}{\kappa}\right)^{3/4}
    d^{-19/8}
    h^{-7/8}
    \omega^{3/2},
\end{equation}
where $\Gamma$ denotes the gamma function.
Using $S(\omega)=N(\omega)|\bm{b}_1\times\bm{b}_2|/(N_x N_y)$ and
$D(\omega)=\partial_\omega N(\omega)/(2N_x N_y)$, we obtain
\begin{equation}
    D(\omega)\simeq
    \frac{\sqrt{3}\Gamma(1/4)}{12\sqrt{2}\pi^{3/2}\Gamma(3/4)}
    \left(\frac{m}{\kappa}\right)^{3/4}
    d^{-3/8}
    h^{-7/8}
    \omega^{1/2}.
    \label{eqmain:analytical_vdos}
\end{equation}

\begin{figure}
    \centering
    \includegraphics[width=0.8\linewidth]{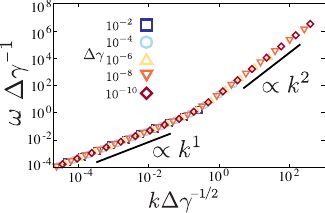}
    \caption{Scaled dispersion relations near yielding for a two-dimensional crystal with WCA interactions for various $\Delta\gamma$, with initial nearest-neighbor distance $2^{1/6} - 0.1$, plotted as $\omega_{-}\Delta\gamma^{-1}$ versus $k\Delta\gamma^{-1/2}$, showing data collapse with linear ($\omega_{-}\sim k$) and quadratic ($\omega_{-}\sim k^2$) regimes and a crossover at $k\Delta\gamma^{-1/2}\sim 1$.}
    \label{figmain:dispersion_wca}
\end{figure}

{\it Generality---}
Figure~\ref{figmain:dispersion_wca} shows the scaled dispersion relations near yielding for a crystal with WCA interactions, $f_{pq,a} = -(\epsilon/r_{pq})
[48(d/r_{pq})^{12}
-24(d/r_{pq})^{6}]
r_{pq,a}\Theta(2^{1/6}d - r_{pq})/r_{pq}$, with units $m=d=\epsilon=1$, confirming the $\omega_{-}\sim k^2$ scaling and data collapse discussed in the main text.

\end{document}